\documentclass[tighten]{aastex62}

\usepackage{booktabs} 
\usepackage{multirow} 
\usepackage{makecell} 

\usepackage{mathtools}

\graphicspath{{./}{figures/}}

\accepted{by ApJL}

\begin{document}

\title{\textbf{Dust resurgence in protoplanetary disks due to planetesimal-planet interactions}}

\correspondingauthor{Lia Marta Bernabò}
\email{lia.bernabo@dlr.de}

\author[0000-0002-8035-1032]{Lia Marta Bernabò}
\affiliation{Department of Physics and Astronomy, University of Padova, Via Marzolo 8, I-35131 Padova, Italy}
\affiliation{Deutsches Zentrum für Luft- und Raumfahrt, Rutherfordstrasse 2, 12489 Berlin, Germany}

\author[0000-0002-1923-7740]{Diego Turrini}
\affiliation{INAF, Osservatorio Astrofisico di Torino, Via Osservatorio 20, I-10025, Pino Torinese, Italy}

\author[0000-0003-1859-3070]{Leonardo Testi}
\affiliation{European Southern Observatory, Karl-Schwarzschild-Str. 2, 85748 Garching bei München, Germany}
\affiliation{INAF – Osservatorio Astrofisico di Arcetri, Largo E. Fermi 5, I-50125 Firenze, Italy}

\author[0000-0003-0724-9987]{Francesco Marzari}
\affiliation{Department of Physics and Astronomy, University of Padova, Via Marzolo 8, I-35131 Padova, Italy}

\author[0000-0002-7657-7418]{Danai Polychroni}
\affiliation{Independent researcher}

\section{Abstract}

Observational data on the dust content of circumstellar disks show that the median dust content in disks around pre-main sequence stars in nearby star forming regions seem to increase from $\sim$1~Myr to $\sim$2~Myr, and then decline with time. 
This behaviour challenges the models where the small dust grains steadily decline by accumulating into larger bodies and drifting inwards on a short timescale ($\le$1~Myr). In this Letter we explore the possibility to reconcile this discrepancy in the framework of a model where the early formation of planets dynamically stirs the nearby planetesimals and causes high energy impacts between them, resulting in the production of second-generation dust. 
We show that the observed dust evolution can be naturally explained by this process within a suite of representative disk-planet architectures.

\section{Introduction}

\textcolor{black}{Dust in protoplanetary disks is expected to rapidly settle on the midplane and grow to pebbles sizes \citep[e.g.][]{natta2007,2014prpl.conf..339T,birnstiel2016}, subsequent growth to planetesimals and planetary cores then occuring through rapid accretion and instabilities \citep[e.g.][]{2014prpl.conf..547J,johansen2017,johansen2019}. In addition to the planet formation route,  
dust can also progressively be removed from the outer regions of disks via gas--induced radial drift  \citep[e.g.][]{weidenschilling1977,birnstiel2016,toci2021}.
ALMA surveys provide growing evidence for rapid depletion of dust and early formation of planetary bodies in disk populations \citep[e.g.][]{Testi2016,Manara2018,2019ApJ...875L...9W}. However, one surprising result of the analysis of the ALMA disks survey data is the apparent lack of decay, and even increase, of the dust content in the first 1-3 Myr of disk evolution \citep[e.g.][]{2019A&A...626A..11C,2019ApJ...875L...9W,Testi2021}. While planet formation-induced substructures in the disk gas distribution may effectively create dust traps and slow down the dust drift, the concentration of dust in traps is expected to promote planetesimal growth and planet formation \citep[e.g.][]{carrera2021,eriksonn2021}. As a result, even in dust traps the amount of dust in disks should steadily decrease over time. \\
\indent The meteoritic constraints from the Solar System also confirm that dust coagulation and the assembly of planetesimals has to occur on fast timescales of the order of 1~Myr or shorter  \citep[e.g.][]{scott2007,nittler2016,wadhwa2020}. The dust content of the disk populations in star-forming regions with different ages should therefore show a rapid and monotonic decline of the dust content  from ages $\le$~1Myr.} However, the homogeneous analysis of ALMA data performed by \cite{Testi2021} appears to suggest that disks in star forming regions with average stellar ages of 1~Myr or less possess a similar or lower median dust content as compared to older 2-3 Myr disks \citep[see also the discussion in][]{2019A&A...626A..11C,2019ApJ...875L...9W}. While these results are still uncertain and rely on the assumptions that the dust properties and conversion factors from observed millimetre flux as well as the dust mass remain constant across disk evolution around pre-main sequence stars \citep[see discussion in][]{Testi2021}, the expected significant \textcolor{black}{monotonic} decrease in median dust mass seems to begin only after 2-3~Myr.

In this work we explore a possible  explanation for this apparently contradictory observational result and show that the observations are indeed consistent with the expectations from planet formation. 
Specifically, as discussed in \cite{Turrini2012}, \cite{Turrini2019} and \cite{2019A&A...629A.116G} the classical view of dust evolution in disks neglects the dynamical and collisional effects of planetesimal-planet interactions. \cite{Turrini2019} showed in the case of the disk around HD\,163296 that the formation of massive planets triggers a phase of strong dynamical excitation of the planetesimal population embedded in the disk, leading to eccentric orbits, high collisional probabilities, and dust production. The apparent rise of the dust median values at around $\sim$2~Myr and the following decrease may thus be a consequence of the planet formation process and its dynamical effects on the disk.

In this Letter we compare the median dust mass estimates from millimeter observations of disks in different star forming regions characterized by various ages with the predicted dust content evolution in disks based on theoretical models that include the formation of second-generation dust produced by planetesimal collisions. \textcolor{black}{In particular, we focus on planets massive enough to dynamically stir the planetesimal disk: this definition roughly encompasses all planets more massive than a few Earth masses independently on the specific type of planets.} Our results illustrate that the observed median dust evolution is well reproduced by models that consider the collisional evolution of planetesimals in systems characterized by a representative set of common planetary architectures. 

\section{Observational data: dust mass vs. age}

The dust mass content in large samples of protoplanetary disks is now available thanks to the observations by the Atacama Large Millimeter/submillimeter Array (ALMA). In this work we focus on the dust mass estimates in disk populating the following star forming regions, ordered by age: Corona Australis region \citep{Cazzoletti2019} aged 0.6 Myr, Taurus region \citep{2019ApJ...872..158A} aged 0.9 Myr,
L1688 region \citep{Williams2019, Testi2016, Testi2021} aged 1.0 Myr, Lupus region \citep{Ansdell2018} aged 2.0 Myr, Chamaeleon region \citep{Pascucci2016} aged 2.8 Myr, and Upper Scorpius region \citep{Barenfeld2016} aged 4.3 Myr. For all these regions we use the compilations of \cite{Testi2021}, which include a homogeneous recalculation of the disk mass and the adoption of GAIA distances and membership analyses. The median ages we use for each of the  stellar populations are also computed in \cite{Testi2021}. Due to the possible dependence of the disk evolution on the initial disk mass and, therefore, on the stellar mass, we focus our analysis on two specific mass groups. 
The first group contains disks around solar type stars with masses in the range  $0.5 M_{\oplus} \leq M_* \leq 1.6  M_{\oplus}$,  while the second group contains disks around red dwarf stars with masses in the range $0.2 M_{\oplus} \leq M_* \leq 0.4  M_{\oplus} $. 



Fig.~\ref{compa} shows the estimated dust masses plotted vs. the median age of the star forming region in the two stellar mass groups. The data in this figure clearly show that the dust content in disks increases between $\sim$0.5-1~Myr and $\sim$2Myr, then declines monotonically only after about 2-3 Myr, instead of the expected fast decay due to the combined effects of grain growth, radial drift, and accretion.

\section{Numerical model of dust rejuvenation}\label{sec:methods}

The n-body simulations are performed with the code {\sc Mercury-Ar$\chi$es} \citep{Turrini2019,Turrini2021}, which allows to model the effects of the mass growth and orbital migration of forming middle-mass and massive planets on the dynamical evolution of planetesimals alngside those of gas drag and the disk self-gravity. The setup of the n-body simulations follows those adopted in \citet{Turrini2019} and \citet{Turrini2021}, with the planetesimal disk extending up to the characteristic radius of the host protoplanetary disk. 
We adopt characteristic radii $r_{c}$ of 30 AU for the disks around red dwarf stars and 50 AU for those around solar type stars and describe the gas surface density with the exponentially tapered power-law $\Sigma(r)=\Sigma_{0} \left( r/r_{c} \right)^\gamma exp\left[-\left(r/r_{c} \right)^{\left( 2-\gamma \right)}\right]$. 
We adopt an exponent $\gamma=0.8$ \citep{Isella2016}, with the gas surface density at the characteristic radius set at 5.9 and 22.7 g/cm$^2$ respectively. This choice results in disks of 0.03 and 0.003 M$_{\odot}$ around solar type and red dwarf stars respectively. The masses of the central stars were set to 0.3 M$_{\odot}$ for red dwarf stars and 1 M$_{\odot}$ for solar type stars. We set the spatial density of massless particles in the n-body simulations to 1000 particles/AU, with the inner edge of the planetesimal disk at 1 AU and the outer edge at $r_{c}$.\\
\indent The damping effects of gas drag on the massless particles are simulated following the treatment from \citet{brasser2007} with updated drag coefficients from \citet{nagasawa2019}. The exciting effects of the disk self-gravity are simulated based on the analytical treatment \textcolor{black}{for axisymmetric disks} by \citet{ward1981} following \citet{marzari2018} and \citet{nagasawa2019} (see \citealt{Turrini2021} for further discussion). \textcolor{black}{The n-body simulations account for the formation of gaps around gas giant planets, as discussed  below, but do not include the effects of non-axisymmetric perturbations, like spiral arms. The disk gas mass does not decline over time and is constant across the n-body simulations except inside the gaps (see below). Since non-axisymmetric perturbations would increase the planetesimal dynamical excitation by promoting higher eccentricity in their orbits \citep[see e.g.][for an illustrative discussion]{marzari2013}, and since the damping effects of gas drag dominate over the disk self-gravity \citep[e.g.][]{nagasawa2019,Turrini2021} and depend on the gas density, our simulations provide a conservative estimate of the dynamical excitation of the planetesimals.} In computing the dynamical effects of the disk gas in the n-body simulations, all massless particles are characterized by a diameter of 100 km \citep[see][]{KlahrSchreiber2016,johansen2017,Turrini2019} and a density  of 1 g/cm$^3$ \citep[see][]{Turrini2019,Turrini2021}.\\
\indent The formation of the planets is modelled over two growth phases \citep{lissauer2009,DAngelo2010,Bitsch2015,johansen2017,johansen2019,DAngelo2021} using the parametric approach from \citet{turrini2011,Turrini2019,Turrini2021}. The first phase is common to all planets, from super-Earths to gas giants, and accounts for their growth by pebble and planetesimal accretion \citep{Bitsch2015,johansen2017,johansen2019}. The planetary mass evolves as 
$ M_{p}(t)=M_{0}+\left( \frac{e}{e-1}\right)\left(M_{1}-M_{0}\right)\left( 1-e^{-t/\tau_{p}} \right)$
where M$_{0}$=0.01 M$_{\oplus}$ is the initial mass of the planetary seed \citep{johansen2017,johansen2019}, M$_{1}$ is the final mass at the end of the first growth phase, and $e$ is the Euler number. M$_{1}$ is set to 30 M$_{\oplus}$ for gas giants (i.e. planets with final mass $>$30 M$_{\oplus}$) while it matches the final planetary mass in all other cases ((i.e. planets with final masses $\le$30 M$_{\oplus}$, see Table \ref{Inputs_table} for the specific values). The constant $\tau_{p}$ is the duration of the first growth phase and is set to 1 Myr for all planets based on observational \citep{Manara2018} and theoretical constraints on the characteristic timescale for pebble accretion \citep{johansen2017,johansen2019}. \\
\indent Gas giants also undergo the second phase of mass growth, accounting for their runaway gas accretion, where their mass evolves as
$ M_{p}(t)=M_{1}+\left( M_{2} - M_{1}\right)\left( 1-e^{-(t-\tau_{p})/\tau_{g}}\right)$
where M$_{2}$ is the final mass of the gas giant planet and $\tau_{g}$ is the e-folding time of the runaway gas accretion process. The value of $\tau_{g}$ is set to 0.1 Myr based on the results of hydrodynamic simulations \citep{lissauer2009,DAngelo2010,DAngelo2021}, meaning that the gas giants reach more than 99\% of their final mass in about 0.5 Myr from the onset of the runaway gas accretion. \textcolor{black}{During the runaway gas accretion process, giant planets form a gap in the disk gas whose width is modelled as $W_{gap} = C\cdot R_{H}$ \citep{Isella2016,marzari2018}, where the numerical proportionality factor $C=8$ is from \citet{Isella2016}. The gas density $\Sigma_{gap}(r)$ inside the gap evolves over time with respect to the local unperturbed gas density $\Sigma(r)$ as $\Sigma_{gap}(r) = \Sigma(r)\cdot \exp{\left[-\left(t-\tau_{p}\right)/\tau_{g}\right]}$ \citep{Turrini2021}.}\\
\indent The migration of the growing planets is modelled over two migration phases based on the migration tracks from \citet{Mordasini2015} following the parametric approach from \citet{Turrini2021}. During the first growth phase, all planets undergo a linear migration regime with drift rate \citep{hahn2005,Turrini2021} 
$\Delta v_{1} = \frac{1}{2}\frac{\Delta a_{1}}{a_{p}}\frac{\Delta t}{\tau_{p}}v_{p}$
where $\Delta t$ is the timestep of the n-body simulation, $\Delta a_{1}$ is the radial displacement during the first growth phase, and $v_{p}$ and $a_{p}$ are the instantaneous planetary orbital velocity and semi-major axis, respectively. During the second growth phase, gas giants undergo a power-law migration regime with drift rate \citep{hahn2005,Turrini2021} 
$    \Delta v_{2} = \frac{1}{2}\frac{\Delta a_{2}}{a_{p}}\frac{\Delta t}{\tau_{g}}\exp^{-\left(t-\tau_{c}\right)/\tau_{g}} v_{p}$
where $\Delta a_{2}$ is the radial displacement during the second growth phase. In the simulations involving migrating planets (runs 7-14, see Table \ref{Inputs_table}) the final semi-major axis of the innermost planet is always set to 0.5 AU. In the case of gas giants, 40\% of the radial displacement occurs during the first growth phase and 60\% during the second growth phase \citep{Turrini2021}. In all other cases, all the radial displacement occurs during the first growth phase.\\
\indent The orbital elements of the massless particles are recorded every 0.1 Myr and the impact probabilities among them are computed by means of the statistical model by \cite{wetherill1967,greenberg1988,farinella1992}, widely used in collisional studies of the asteroid belt \citep{obrien2011}. The collisional frequencies and the dust production rates over each 0.1~Myr-long time interval are estimated following the approach by \citet{Turrini2019} based on the scaling law for the collisional mass erosion by \citet{Genda2017}. We assume that 20\% of the eroded mass is eventually converted into dust by the collisional cascade due to the increased dust production efficiency of smaller impactors \citep[\citealt{okeefe1985,koschny2001}, see][for discussion]{Turrini2019}. The dust production is computed with the parallel code {\sc Debris} using a particle-based approach instead of the original grid-based approach by \citet{Turrini2019}, as the latter is not accurate in presence of migrating planets. In computing the dust production, each massless particle in the n-body simulations is treated as a swarm of planetesimals.\\
\indent 
The initial mass of each swarm is computed integrating the disk gas density profile over a ring wide 0.1 au centered on the initial orbit of the massless particle, and assuming a dust-to-gas ratio of 0.01  \citep[e.g.][]{bohlin1978,natta2007,ercolano2017}. This results in disks initially containing 110 and 10 M$_{\oplus}$ of dust around solar type and red dwarf stars respectively. The bulk of the initial dust population is assumed to be converted into planetesimals by the time the disks become dynamically excited, as discussed below. 
The size-frequency distribution of the planetesimals in each swarm is that characteristic of a population in a collisional steady-state \citep[ \citealt{Weidenschilling2008,Weidenschilling2011} see][for a discussion]{Turrini2019}, with diameters spanning between 1 m and 400 km \citep{Krivov2018,Turrini2019}.
The total mass of each swam decreases over time from its initial value due to the mass loss associated with the collisional dust production. This approach allows to capture the self-regulating nature of the collisional process under study without having to fully resolve the collisional cascade \citep{Turrini2019}: as a result,  
the more intense the dust production, the faster the depletion of the population of excited planetesimals, which in turn implies a shorter duration of the collisional dust production phase.

\begin{table}[h]
\centering
\begin{tabular}{c | c c c c c c c}
\toprule
Scenario & Stellar & Characteristic & Initial Mass & & Planetary & semi-major \\
& mass & radius \textit{R$_{c}$} & of Dust & & mass & axis\\
\midrule
\textbf{Run 1} & \multirow{4}{1.5cm}{1 M\textsubscript{$\odot$}} & \multirow{4}{1cm}{50 AU} & \multirow{4}{1.5cm}{100 M\textsubscript{$\Earth$}} & & \multirow{2}{1.5cm}{1 M\textsubscript{Jup}} & 5 AU, 11 AU \\
\textbf{Run 2} & & & & & & 5 AU, 11 AU, 22 AU \\ \cline{1-1}
\cline{5-7}
\textbf{Run 3} & & & & & \multirow{2}{1.5cm}{150 M\textsubscript{$\Earth$}} & 5 AU, 11 AU\\
\textbf{Run 4} & & & & & & 5 AU, 11 AU, 22 AU \\

\midrule
\textbf{Run 5} & \multirow{2}{1.5cm}{0.3 M\textsubscript{$\odot$}} &  \multirow{2}{1cm}{30 AU} & \multirow{2}{1.5cm}{10 M\textsubscript{$\Earth$}} & & \multirow{1}{1.5cm}{30 M\textsubscript{$\Earth$}} & 5 AU 8 AU \\
\textbf{Run 6} & & & & & \multirow{1}{1.5cm}{10 M\textsubscript{$\Earth$}} & 5 AU, 8 AU, 11 AU \\
\midrule
\midrule
 &  &  &  & & & Initial semi-& Final semi-\\
&  & & & & & major axis & major axis\\
\midrule
\textbf{Run 7} & \multirow{3}{1.5cm}{1 M\textsubscript{$\odot$}} & \multirow{3}{1cm}{50 AU} & \multirow{3}{1.5cm}{100 M\textsubscript{$\Earth$}} & & \multirow{3}{1cm}{1 M\textsubscript{Jup}} & 5 AU & \multirow{3}{1cm}{0.5 AU}\\
\textbf{Run 8} & & & & & & 11 AU & \\
\textbf{Run 9} & & & & & & 22 AU & \\
\hline
\textbf{Run 10} & \multirow{5}{1.5cm}{0.3 M\textsubscript{$\odot$}} & \multirow{5}{1cm}{30 AU} & \multirow{5}{1cm}{10 M\textsubscript{$\Earth$}} & & \multirow{3}{1cm}{30 M\textsubscript{$\Earth$}} & 11 AU & \multirow{5}{1cm}{0.5 AU}\\
\textbf{Run 11} & & & & & & 8 AU, 11 AU &\\
\textbf{Run 12} & & & & & & 8 AU &\\ \cline{1-1}
\cline{5-7}
\textbf{Run 13} & & & & & \multirow{2}{1cm}{10 M\textsubscript{$\Earth$}} & 11 AU & \\
\textbf{Run 14} & & & & & & 8 AU, 11 AU & \\

\bottomrule
\end{tabular}
\caption{The set or representative disk-planet architectures used in the simulations of this study. The values reported are those at the beginning of the simulations, which corresponds to 0.2 Myr old disks.
When multiple initial semi-major axis are reported, their number matches the number of massive planets considered in that specific scenario. When a final semi-major axis is specified (runs 7-14), the simulation includes orbital migration otherwise the planets are assumed to form \textit{in situ} (runs 1-6).} 
\label{Inputs_table}
\end{table}

\begin{figure*}
    \centering

   \includegraphics[width=\textwidth]{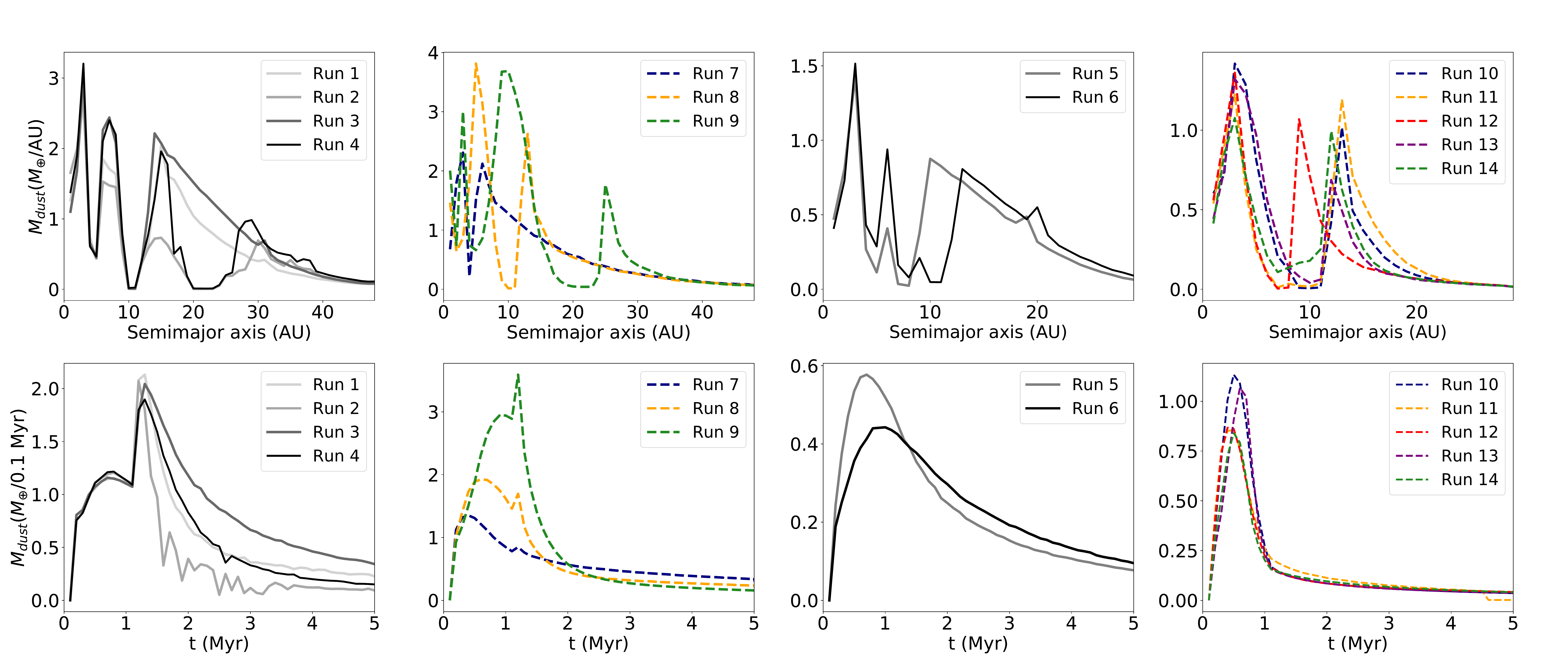}
   \caption{
   \textcolor{black}{The plots show the dust production across the radial extension of the disk (in M$_\oplus$ per 1 AU-wide ring integrated over time, top row) and across time (in M$_\oplus$ produced every 0.1 Myr spatially integrated over the whole disk, bottom row) for all runs. In both rows the plots show the dust production for planets forming \textit{in situ} (first and third panels) and for migrating planets (second and fourth panels). The first two panels in each row refer to solar type stars, the last two to red dwarf stars.}}
   \label{Run6}

\end{figure*}

\textcolor{black}{The n-body simulations start from the time the 0.01~M$_\oplus$ seeds appear in the disks. To compare the simulated systems with the observed data, we assume that such planetary seeds form in t$_0$=0.5~Myr. By 1~Myr in the timescale of the observed systems (hence, 0.5~Myr in the timescale of the simulations), the simulated disks will contain planets at least a few M$_\oplus$ in mass, more than 90\% of the initial dust population will be incorporated into planetesimals (see the curves labelled \textit{planetesimals formation} in Fig. \ref{compa} \textcolor{black}{and Sect. \ref{sec:results} for further discussion}), and the planetesimals will start being dynamically excited by the planets (see Fig. \ref{Run6}). The dust produced over each 0.1~Myr-long interval is added to the existing dust population, and the resulting total dust declines with time following a power law with characteristic timescale of $\simeq$1.5~Myr to account for the continuing dust evolution and the decreasing efficiency of planetesimal formation over time \citep[see][for a discussion]{johansen2019}.} The total dust mass evolution with time can then be compared with the observed data.\\
\textcolor{black}{\indent The set of scenarios considered for both solar type and red dwarf stars, summarized in Table 1, encompasses architectures with two or three ``cold'' planets forming \textit{in situ} as well as with one or two planets migrating across the disk from different starting positions to become ``hot'' planets. This choice is motivated by the goal of comparing the median dust evolution of the observed disk populations with a set of planet-forming scenarios that are representative, within the limits of our current observational capabilities, of common planetary architectures among exoplanets. As a result, we focused our exploration on planetary systems hosting 1-3 planets: although this may be an observational bias, single-planet systems currently account for the majority of known exoplanetary system (source: The Extrasolar Planets Encyclopaedia\footnote{\url{http://.exoplanet.eu}}), while planetary systems hosting 2-3 planets dominate the known population of multi-planet systems \citep{ZinziTurrini2017,Turrini2020}. Furthermore, the semi-major axes of the migrating and non-migrating planets we considered span the orbital range ($\sim$0.1-20 AU) where the vast majority of the currently confirmed exoplanets reside (source: NASA Exoplanet Archive\footnote{\url{https://exoplanetarchive.ipac.caltech.edu}}).}


%

\section{Comparison between model results and observations}\label{sec:results}

\begin{figure}
\includegraphics[width=0.5\columnwidth]{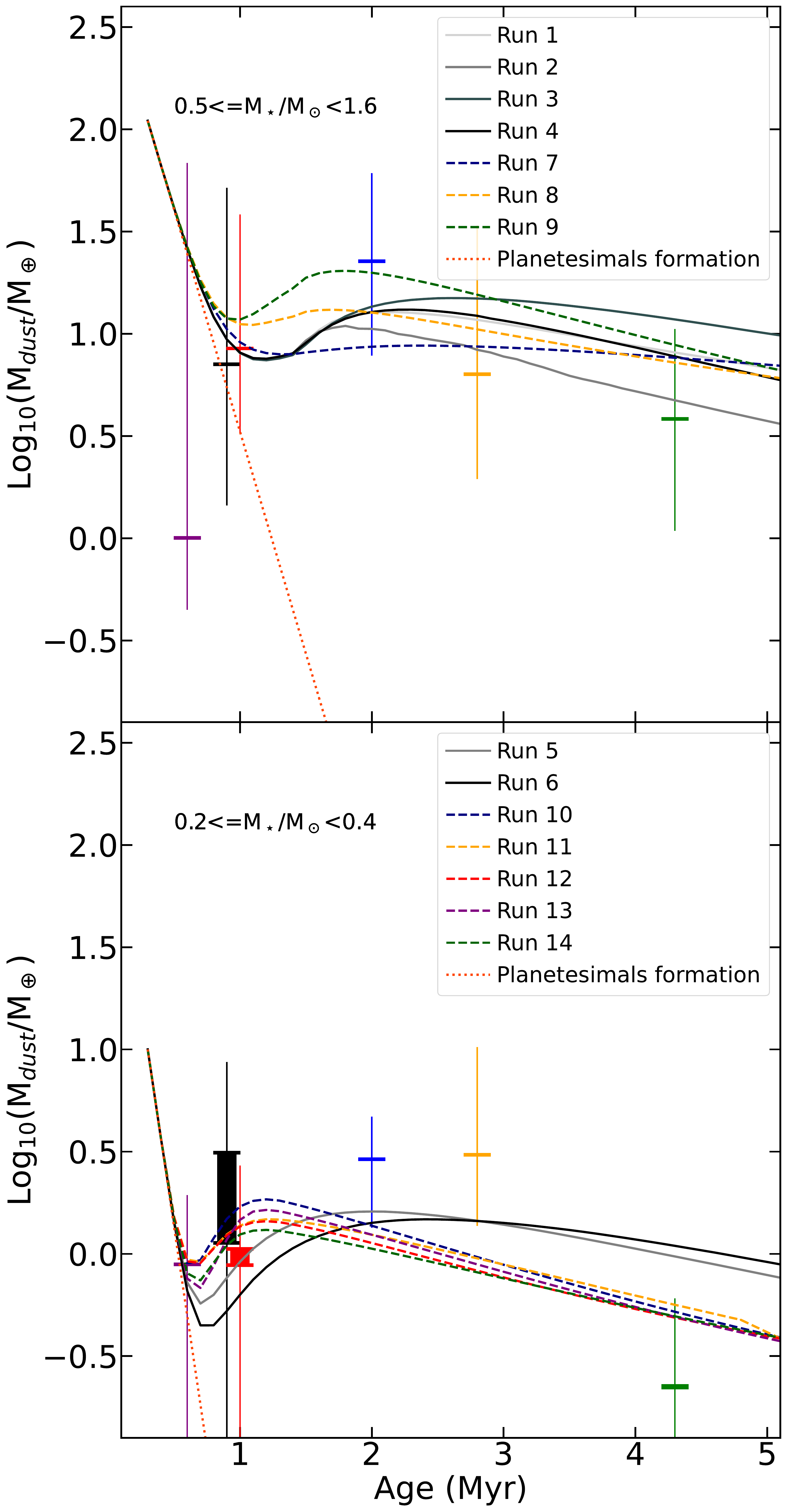}
\caption{Dust mass vs age of the star-forming region for disks around solar type stars (\textit{top panel}) and red dwarf stars (\textit{bottom panel}). For each star forming region the \textcolor{black}{thin vertical lines} highlight the 25\% and 75\% percentiles, while the horizontal line marks the median value. \textcolor{black}{The thick vertical bars, where present, show the possible range of median values \citep[see ][for a detailed explanation on how the range is computed]{Testi2021}. The red dotted line shows the expected dust decrease due to the formation of planetesimals when no dust rejuvenation is included.}} 
\label{compa}
\end{figure}

\textcolor{black}{Fig.~\ref{Run6} illustrates the collisional dust production in the simulated planet-forming scenarios reported in Table \ref{Inputs_table} plotted as a function of the radial distance (top panels) and time (bottom panels).} In Fig.~\ref{compa} we compare the dust observed in the considered disk population with the predictions of our simulations of dust rejuvenation. The simulated dust production scenarios \textcolor{black}{are capable of successfully reproducing} the re-growth and the slower decline of the dust content revealed by the observational data. The initial rapid decline of dust content before 1 Myr\textcolor{black}{, due to the dust growth into planetesimals, is followed by a rise in dust production linked to the appearance of the planets and the high-velocity collisions among planetesimals they create. The secular evolution of the dust,  combined with the decline of the population of excited planetesimals due to collisional erosion and gas drag, naturally explains the secular, slow decay of the dust mass content with time observed in the considered disk populations.} 

\textcolor{black}{As illustrated by the top panels of Fig. \ref{Run6} the dust rejuvenation process can affect a significant fraction of the radial extension of circumstellar disks. In the simulated scenarios and for the assumed dust production efficiency, planetesimal impacts can convert between 25\% and 50\% of the mass of the planetesimal disk back into dust. As dust production linearly depends on the dust production efficiency and the mass of the planetesimal disk, our results can be readily scaled to different values of the latter quantities. \textcolor{black}{As discussed in Sect. \ref{sec:methods}, our scenarios assume an efficient conversion 
of the initial dust budget into planetesimals to fit the observational data. Lower planetesimal formation efficiencies would proportionally reduce the collisional production of dust and the overall dust population at later ages, and would result in disks containing varying mixtures of primordial and second-generation dust \citep{Turrini2019}. Disks characterized by very low planetesimal formation efficiency (e.g. $\sim$10\%) would result in limited collisional dust production. However such scenario would be at odds with the observed dust masses of disks between 1 and 2 Myr \citep{Testi2021}, as well as with the masses of known exoplanetary systems \citep{Testi2016,Manara2018,mulders2021} and the metallicities of known giant planets \citep{Thorngren2016,Shibata2020,Turrini2021,turrini2021b}.}
The duration of the dust rejuvenation process in our simulations ranges between 1 and 2~Myr, depending on the formation history of the system (see Fig. \ref{Run6}, bottom panels). This duration matches well with the age interval of observed disks with rising dust population (see Fig. \ref{compa}). As illustrated by the presence of two peaks in the dust production over time due to forming gas giants (see in particular the bottom left panel of Fig. \ref{Run6}), the dust rejuvenation process is triggered already when planets of a few M$_\oplus$ in mass are present within circumstellar disks, although the dust production is less intense than when giant planets appear.}

\textcolor{black}{As shown by Figs. \ref{Run6} and \ref{compa}, the spatial distribution and temporal evolution of the dust rejuvenation process strongly depend on the architecture and formation history of the simulated systems. Circumstellar disks hosting significantly less massive planets than those here considered may be affected by limited dust rejuvenation, making their dust population indistinguishable from that of disks that host no planet. Alternatively, their dust rejuvenation could be characterized by a short peak duration followed by a rapid decline, resulting in curves crossing the low-end tails of the dust distributions observed in Chamaeleon and Upper Scorpius (see Fig. \ref{compa}). Conversely, systems characterized by more extreme migration histories of gas giants (as recently suggested for the Solar System by \citealt{pirani2019,oberg2019}, and V1298 Tau by \citealt{mascareno2021}) or higher multiplicities of migrating planets (e.g. Trappist-1 in the case of red dwarf stars, \citealt{tamayo2017,papaloizou2018}, and the Solar System, \citealt{pirani2019}) could be characterized by stronger dust rejuvenation processes and produce dust populations consistent with the high-end values observed in Lupus and Chamaeleon (see Fig. \ref{compa}).}

\section{Conclusions}

The fast decay of dust mass in circumstellar disks expected on the basis of grain growth and radial drift is not observed in the surveys of circumstellar disks of different ages by ALMA. 
A possible explanation for the delayed disappearance of the dust is the formation of second-generation dust by planetesimal collisions excited by the formation of \textcolor{black}{planets more massive than a few M$_{\Earth}$, the specific mass values depending on the architecture of individual systems.} 
The models we presented for the rejuvenation of the dust content in disks closely match the observations and provide a strong support to the scenario where the formation of \textcolor{black}{planets 
can lead} to a resurgence of the dust mass. This process increases the lifetime of the dust component of circumstellar disks, delaying the final disappearance of the dust to after the formation of the planets. Within this scenario, the observational data on the dust abundances of disks in star-forming regions of different ages indicates that the timescale for the complete formation of planets can be directly constrained by the peaks in the dust abundances to $\le$1-2 Myr. Finally, the phase of intense planetesimal collisions linked to the formation of planets and the associated collisional erosion of the larger planetesimals can provide a natural mechanism to produce the large population of small planetesimals invoked by \cite{krivov2021} to explain the estimated dust masses in the population of debris disks.

\acknowledgments
This work was partly supported by the Italian Ministero dell'Istruzione, Universit\`a e Ricerca through the grant Progetti Premiali 2012 – iALMA (CUP C$52$I$13000140001$), by the Deutsche Forschungs-gemeinschaft (DFG, German Research Foundation) - Ref no. 325594231 FOR $2634$/$1$ TE $1024$/$1$-$1$, by the DFG cluster of excellence Origins (www.origins-cluster.de), and by the Italian National Institute of Astrophysics (INAF) through the projects PRIN-INAF 2016 ``The Cradle of Life - GENESIS-SKA (General Conditions in Early Planetary Systems for the rise of life with SKA)'', PRIN-INAF 2019 ``Planetary systems at young ages (PLATEA)'', and the Main Stream projects ``Ariel and the astrochemical link between circumstellar discs and planets'' (CUP: C54I19000700005) and ``Non-spherical dust dynamics in protoplanetary disks: how dust particle realistic shapes change the dust evolution timescales'' (CUP: C54I19000460005). This project has received funding from the European Union's Horizon 2020 research and innovation programme under the Marie Sklodowska-Curie grant agreement No 823823 (DUSTBUSTERS) and from the European Research Council (ERC) via the ERC Synergy Grant {\em ECOGAL} (grant 855130).



\setlength{\tabcolsep}{8.0pt} 

\typeout{}
\bibliography{letter}

\end{document}